\documentclass[twocolumn,aps,prl,superscriptaddress,showpacs]{revtex4-1}
\usepackage[utf8]{inputenc}
\usepackage{color}
\usepackage{amsmath}
\usepackage{amssymb}
\usepackage{graphicx}
\usepackage{esint}
\usepackage[unicode=true,
 bookmarks=false,
 breaklinks=true,pdfborder={0 0 0},backref=false,colorlinks=true]
 {hyperref}
 \usepackage{float}
\usepackage[title]{appendix}

\makeatletter

\usepackage{epstopdf}\usepackage{subfigure}
\usepackage{float}
\usepackage{amsfonts}
\usepackage{bm}
\usepackage{subfigure}
\usepackage{braket}

\makeatother
\usepackage{tikz}

\begin{document}

\title{Chiral phonons and pseudo-angular momentum in non-symmorphic systems}

\author{Tiantian Zhang}
\affiliation{Department of Physics, Tokyo Institute of Technology, Ookayama, Meguro-ku, Tokyo 152-8551, Japan}
\affiliation{Tokodai Institute for Element Strategy, Tokyo Institute of Technology, Nagatsuta, Midori-ku, Yokohama, Kanagawa 226-8503, Japan}
\author{Shuichi Murakami}
\affiliation{Department of Physics, Tokyo Institute of Technology, Ookayama, Meguro-ku, Tokyo 152-8551, Japan}
\affiliation{Tokodai Institute for Element Strategy, Tokyo Institute of Technology, Nagatsuta, Midori-ku, Yokohama, Kanagawa 226-8503, Japan}
\begin{abstract}
Chiral phonons are the ones with nonzero polarization and can be observed only via a selective coupling with valley electrons and circularly polarized photons. In such process, a new physical quantity, i.e., pseudo-angular momentum (PAM), is required to meet the selection rule. 
However, phonon PAM was thought to be quantized and can be only defined in the symmorphic systems. 
In this work, we generalized the definition of PAM to three-dimensional non-symmorphic systems, which show distinct different properties compared with the one in symmorphic systems, $e.g.$, PAM can be non-quantized and $\mathbf{q}$-dependent  but still be an observable quantity by experiments. Such new definition and discoveries can help us to obtain chiral phonons in a broader class of systems with a nonzero group velocity and to convey information like chirality and angular momentum in solids as expected. Materials are also offered to understand the new definition and for further experimental detection. 


\end{abstract}

\maketitle

\paragraph*{Introduction}

Chirality has been widely studied in fundamental physics since it reveals symmetry breaking of (quasi-)particles and governs many unconventional phenomena, such as chiral anomaly \cite{nielsen1983adler,son2013chiral,huang2015observation}, unconventional Landau spectrum \cite{zhang2005experimental,zhao2021index}, Klein tunneling \cite{katsnelson2006chiral}, nontrivial surface states \cite{zhang2018double,lu2014topological,lu2015experimental,weng2015weyl,Weyl_newfermions,Weyl_Taas,Weyl_experiments,Weyl_exp,Weyl_phon_mecha,Weyl_acoustic3,Weyl_acoustic4} and so on. In the recent years, chiral phonons draw much attention due to their contribution to many physical processes, like electronic phase transition \cite{rini2007control,forst2015mode}, giant thermal Hall effects in cuprate high-$T_{c}$ superconductors \cite{grissonnanche2020chiral,grissonnanche2019giant} and valley polarization \cite{zeng2012valley,carvalho2017intervalley,cao2012valley,wang2012electronics,he2014tightly,wu2013electrical,jones2013optical,xu2014spin}. 
Chiral phonons can also transport information like chirality and angular momentum if they further couple with valley electrons and photons. 
However, most of them were only studied in two-dimensional (2D) systems at either the Brillouin zone (BZ) center or its boundaries \cite{chen2021probing,drapcho2017apparent,tatsumi2018interplay,malard2009raman,zhu2018observation,chen2019entanglement,li2019momentum,li2019emerging}, which means that they will have a vanishing group velocity due to the local extremal energy and reduces the efficiency of the information propagation. 
Thus, studies on chiral phonons in three-dimensional (3D) systems with a non-vanishing group velocity will help us to realize the chiral phonon-based quantum devices through materials.

In the previous studies, pseudo-angular momentum (PAM) is proposed as a quantized physical quantity defined by the eigenvalue of a rotation operator \cite{zhang2015chiral} to offer a crucial selection rule to study chiral phonon absorption/emission in the intra/inter-valley electron scattering process via photon absorption. 
Although chiral phonons at high-symmetry points (HSP) were observed by the helicity-resolved Raman scattering experiments in transition-metal dichalcogenides (TMD) and graphene by detecting the helicity flipping of the incident and emission photons \cite{zhu2018observation,chen2019entanglement,hanbicki2018double,kunstmann2018momentum,ferreira2010evolution,hao2010probing,eckmann2012probing,ni2008raman}, such a definition restricts the study of chiral phonons only at high-symmetry points/lines (HSPs/HSLs) with local rotation symmetry in symmorphic space groups \cite{chen2021propagating}. 
{Chiral phonons and phonon PAM have never been discussed in non-symmorphic space groups, in which PAM was thought to be ill defined. 
In this work, we propose that PAM can be non-quantized and $\mathbf{q}$-dependent at any momenta with (screw) rotation symmetry in the BZ, and we generalize the definition to 3D space groups, which will be much easier for us to obtain chiral phonons with nonzero group velocity to convey information along expected directions. }

\begin{center}
\begin{figure}
\includegraphics[scale=0.86]{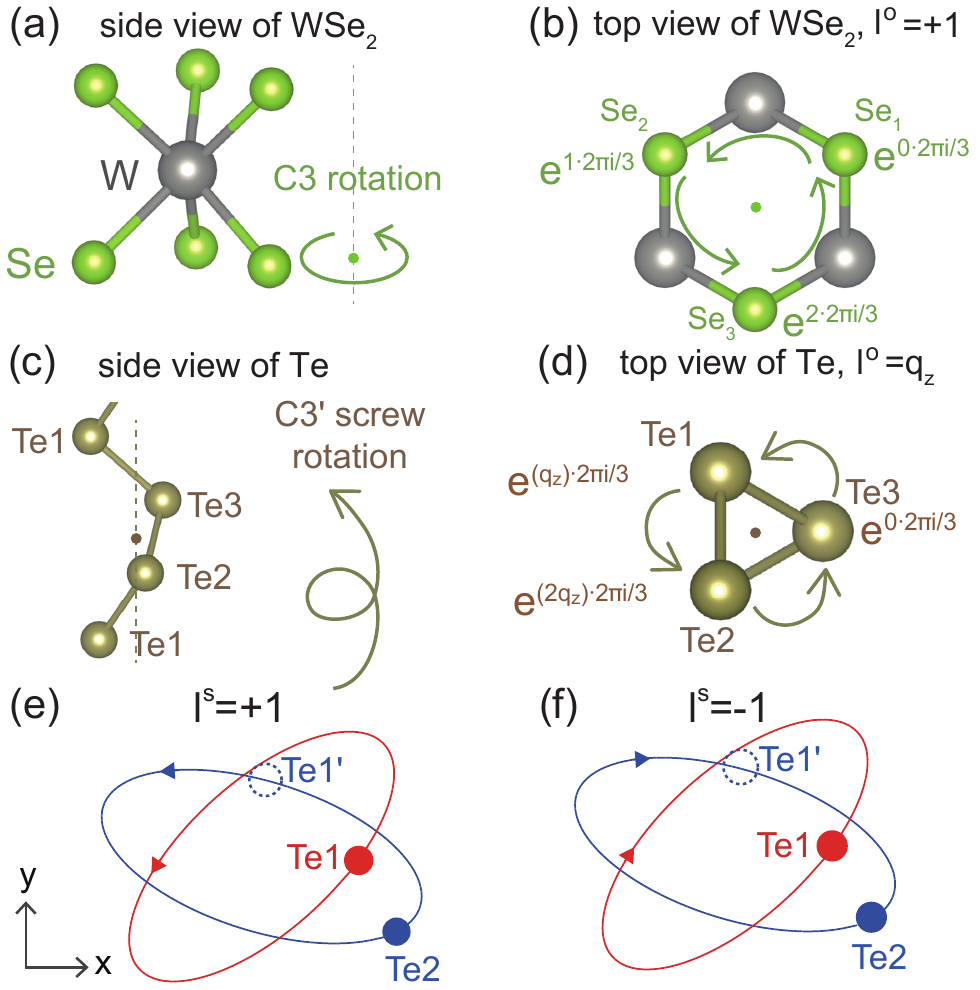}\caption{(a-b) Side and top view of 2-dimensional material WSe$_{2}$ with three-fold rotation symmetry $\hat{C_{3}}$. (b) also shows the phase change of the phonon nonlocal part for selenium at $K$ ($\frac{1}{3},\frac{1}{3},0$) under $\hat{C_{3}}$, which corresponds to an orbital pseudo-angular momentum of $l^{o}$=+1. (c-d) Side view and top view of 3-dimensional material tellurium with screw three-fold rotation symmetry $\hat{C_{3}^{\prime}}$. (d) also shows the phase change of the phonon nonlocal part for tellurium at ($\frac{1}{3},\frac{1}{3},q_{z}$) under $\hat{C_{3}^{\prime}}$, which corresponds to an orbital pseudo-angular momentum of $l^{o}=+q_{z}$. (e-f) Schematic figures for two different vibration trajectories of the third and first modes for tellurium along K-H high-symmetry line ($\frac{1}{3},\frac{1}{3},\frac{1}{4}$), which have a spin pseudo-angular momentum of $l^{s}$=$+1$, $-1$, respectively.\label{fig:ScrewRot}}
\end{figure}
\end{center}

\paragraph{Chiral phonons and phonon polarization}
{The chirality of phonons can be understood from the phonon polarization point of view, and regarded as one of the information carried by phonons. }
In 2D materials, phonon circular polarization for each mode is along the out-of-plane direction, while in 3D materials it can point to any directions due to the interatomic coupling in all directions. For example, the vibration trajectories for selenium atoms in WSe$_{2}$ in Fig.~\ref{fig:ScrewRot} (a-b) are in a circular shape in the $xy$-plane due to the 2D nature of WSe$_{2}$, while it is in an elliptical shape in the $xy$-plane for the tellurium atoms in elemental tellurium as shown in Fig.~\ref{fig:ScrewRot} (e-f). 
Yet, we can still define the phonon polarization, i.e., phonon chirality, for a 3D system by the vibration trajectory of atoms in the real space. 
{For example, if the atom vibrates in a counterclockwise circular/elliptical trajectory on a 2D plane, it will carry a counterclockwise chirality labelled by ``$+$'', as shown in Fig. \ref{fig:ScrewRot} (e); if the atom vibrates in a clockwise trajectory, it will carry a clockwise chirality labelled by ``$-$'', as shown in Fig. \ref{fig:ScrewRot} (f). }


\paragraph{Redefinition for PAM in non-symmporphic systems}
Chiral phonons play an important role in the phonon-involving optical transition of the valley electron scattering process. For example, 
the intra-valley electron scattering may involve chiral phonons at the BZ center, while the inter-valley one may involve chiral phonons at any momentum  in the BZ, and it can be probed by photons with different circular polarization. 
As a selection rule for such optical transition process, the PAM offers an additional conservation condition besides the crystal momentum conservation and energy conservation in the valley-dependent optoelectronics:

\begin{equation}
l_{e}^{v}+m=l_{e}^{c}+l_{ph},
\end{equation}
\begin{equation}
\hbar\mathbf{k}_{e}^{v}=\hbar\mathbf{k}_{e}^{c}+\hbar\mathbf{q}_{ph},
\end{equation} 
\begin{equation}
E_{e}^{v}+E_{photon}=E_{e}^{c}+E_{ph},
\end{equation} 
where $l_{e}^{v/c}$ is the electron PAM for the valence/conduction band, $l_{ph}$ is the PAM for a phonon, $m$ = $\pm1$ represent the right/left-circularly polarized light, $\mathbf{k}_{e}$ and $\mathbf{q}_{ph}$ are the crystal momentum for electrons and phonons, $E_{e/photon/ph}$ is the energy for electrons/photons/phonons.

In the previous studies on chiral phonons, the PAM was thought well defined only at HSPs and along HSLs in symmorphic space groups, in terms of the eigenvalue of $n$-fold rotation symmetry :
\begin{equation}
\hat{C_{n}}u_{\mathbf{q}}=e^{\frac{-2\pi i}{n}\cdot l_{ph,\mathbf{q}}}u_{\mathbf{q}},
\label{eq:4}
\end{equation}
where $\hat{C_{n}}$ is the $n$-fold rotation operator, $u_{\mathbf{q}}$ is the phonon Bloch wave function, and $l_{ph,\mathbf{q}}$ is the phonon PAM at momentum ${\mathbf{q}}$. Thus, in order to have a well defined $l_{ph,\mathbf{q}}$, which is quantized with discrete values of \{0, 1, ... $n$-1\} $mod$ $n$, chiral phonons can be only studied at rotation-preserving momenta in symmorphic space groups. 

However, in the inter-valley electron scattering via photon absorption, the electronic valleys can exist at any momenta in the BZ, which means that the momentum of the chiral phonon involved in the optical transition process can take any values. Moreover, materials with valleys at non-HSPs are very common in systems with (screw) rotation symmetry \cite{PhysRevB.92.115202,PhysRevB.94.035304,PhysRevB.92.085406}. Thus, it is necessary for us to have a universal definition for the PAM in both electronic and phononic systems, especially in non-symmorphic systems. 

In this paper, we propose that for non-symmorphic systems with screw rotation symmetry, $\hat{C_{n}^{\prime}}$, the phonon (electron) PAM is still well defined and $\mathbf{q}$-dependent ($\mathbf{k}$-dependent for electrons) along the $\hat{C_{n}^{\prime}}$-preserving HSL via Eq. (\ref{eq:4}) with $\hat{C_{n}}$ replaced by $\hat{C_{n}^{\prime}}$, 
but not quantized due to the fractional translation of the non-symmorphic symmetry. For example, Te with a non-symmorphic space group $P3_{1}21$ has a threefold screw rotation symmetry $\hat{C_{3z}^{\prime}}$ along $z$-direction. $l_{ph,\mathbf{q}}$ for each phonon mode of Te is also defined as the eigenvalue of $\hat{C_{3z}^{\prime}}$ and $\mathbf{q}$-dependent along $\Gamma$-$Z$ and $H$-$K$ HSLs, i.e., $l_{ph,\mathbf{q}}=1+q_{z}, q_{z}, q_{z}-1$ $mod$ 3 with $q_{z}\in[-0.5,0.5]$. 
The definition can be extended for any (quasi-)particles at any momenta in solids preserving (screw) rotation symmetry.

\paragraph{Definition for orbital and spin part of PAM}
Since the phonon Bloch wave function $u_{\mathbf{q}}$ has both a nonlocal contribution (intercell part) from the Bloch phase factor $e^{i\cdot \mathbf{R}_{l} \cdot \mathbf{q}}$ ($\mathbf{R}_{l}$ is the atomic position) and a local contribution (intracell part) from relative vibrations between sublattices, 
phonon PAM can be decomposed into the orbital part $l^{o}$ and the spin part $l^{s}$, i.e., $l^{ph}=l^{s}+l^{o}$, which will offer a new way to understand phonon PAM.  
Both $l^{o}$ and $l^{s}$ should be defined along the (screw) rotation axis direction since the total PAM $l^{ph}$ is defined by the (screw) rotation symmetry.  

The orbital part of PAM, i.e., $l^{o}$, is the phase difference of the Bloch phase factor between atoms in different sublattices related by (screw) rotation symmetry, and it depends on the atomic sites $\mathbf{R}_{l}$, phonon momentum $\mathbf{q}$ and phonon modes. 
$l^{o}$ is a quantized integer with values of 0, $\pm1$ for symmorphic systems. 
For example, Fig. \ref{fig:ScrewRot} (b) shows the phases of the phonon nonlocal part $e^{i\cdot \mathbf{R}_{l} \cdot \mathbf{q}}$ at $\mathbf{q}$ = $K$ ($\frac{1}{3},\frac{1}{3},0$), and the fractional atomic positions for the three $\hat{C_{3z}}$-related selenium atoms in WSe$_{2}$ are $\mathbf{R_{1}}$ = ($\frac{1}{3},-\frac{1}{3},\frac{1}{2}$), $\mathbf{R_{2}}$ = ($-\frac{2}{3},-\frac{1}{3},\frac{1}{2}$) and $\mathbf{R_{3}}$ = ($\frac{1}{3},\frac{2}{3},\frac{1}{2}$) for Se$_{1}$, Se$_{2}$ and Se$_{3}$ in Fig. \ref{fig:ScrewRot} (b). 
Thus, $e^{-i\frac{2\pi}{3}l^{o}}$ = $e^{i\cdot\mathbf{q}\cdot(\mathbf{R_{3}-R_{2}})}$ = $e^{i\cdot\mathbf{q}\cdot(\mathbf{R_{2}-R_{1}})}$ = $e^{i\cdot\mathbf{q}\cdot(\mathbf{R_{1}-R_{3}})}$ gives rise to $l^{o}$ = 
+1 for selenium and all the phonon modes in WSe$_{2}$ according to the phase difference. 
However, $l^{o}$ becomes a $\mathbf{q}$-dependent quantity for systems with screw rotation symmetry due to the fractional translation. 
For example, Fig. \ref{fig:ScrewRot} (d) show the phases of phonon nonlocal part $e^{i\cdot \mathbf{R}_{l} \cdot \mathbf{q}}$ at $\mathbf{q}$ = ($\frac{1}{3},\frac{1}{3},q_{z}$), and the fractional atomic positions for the three $\hat{C_{3z}^{\prime}}$-related selenium atoms are $\mathbf{R_{1}}$ = ($x,x,0$), $\mathbf{R_{2}}$ = ($-x,0,\frac{1}{3}$) and $\mathbf{R_{3}}$ = ($0,-x,\frac{2}{3}$) in tellurium. Thus, $(e^{-i\frac{2\pi}{3}l^{o}})^{3}$ = $e^{-i\cdot\mathbf{q}\cdot\mathbf{R_{0}}}$ = $e^{-2\pi i\cdot q_{z}}$, i.e., $l^{o}$ = $q_{z}$ for all the phonon modes in Te according to the phase difference, where $\mathbf{R_{0}}$ is the unit lattice vector along the $c$-axis. 

The spin part of PAM, $l^{s}$, which is a local contribution (intracell part) from the sub-lattice relative vibration, depends on the in-plane atomic vibration. 
$l^{s}$ can directly correspond to the phonon polarization for atoms on a rotation axis in 2D symmorphic systems, i.e., phonon modes with a counterclockwise/clockwise trajectory will give rise to $l^{s}=\pm1$. 
However, in 3D non-symmorphic systems, $l^{s}$ cannot be obtained directly from the phonon polarization or the phonon chirality, but we need to calculate the phase difference between atoms related by screw rotation symmetry. 
For example, Fig. ~\ref{fig:ScrewRot} (e) and (f) show the vibration trajectories of the third and first phonon bands, respectively, for two $\hat{C_{3}'}$-related tellurium atoms at the middle point of $K$-$H$ line ($\frac{1}{3}, \frac{1}{3}, \frac{1}{4}$), where both of those two tellurium atoms have a counterclockwise polarization. Here, in this case, Te$_{1}^{\prime}$, which is the obtained from Te$_{1}$ through the screw rotation $\hat{C_{3}'}$, has a $\frac{2\pi}{3}$ phase advancing to Te$_{2}$ and therefore this case corresponds to $l^{s}$ = $+1$ for Te. 
Likewise, we obtain $l^{s}$ = $-1$ for Te in Fig. ~\ref{fig:ScrewRot} (f), since Te$_{1}^{\prime}$ has a $\frac{2\pi}{3}$ phase behind Te$_{2}$. 
We note that the trajectory of phonon modes for 3D materials may be in an elliptic shape in the 2D plane since they are combinations of linearly vibrations along different directions, but $l^{s}$ is still well defined and quantized. 

\paragraph{Distinct features of PAM in non-symmorphic systems}
We note that the definition of PAM in non-symmorphic systems is not just an extension of that in symmorphic systems, but has unique features: 
(i) By comparing two methods of calculating phonon PAM above, we can show that the non-quantized nature of $l_{ph}$ is from the fractional-translation part of the screw rotation symmetry, which is unique for non-symmorphic systems. Namely, the phase difference between sub-lattices under screw rotation symmetry leads to a non-quantized value of the orbital part of the phonon PAM, $l^{o}$. 
(ii) Another distinct point for PAM in non-symmorphic systems is that we only need to consider the atomic position components along the fractional translation direction of the screw rotation operator when calculating $l^{o}$, which means that the atomic positions in-plane will make no influence on $l^{o}$. 
It is because atoms related by screw rotation symmetry in non-symmorphic systems are different atoms in the same unit cell, while it is not necessarily the case in symmorphic systems as we have shown in WSe$_{2}$. 
Thus, we only need to consider the positions along screw rotation axis in non-symmorphic systems, since the in-plane positions will always contribute zero phase difference in the process of calculating $l^{o}$.

\begin{center}
\begin{figure}
\includegraphics[scale=0.8]{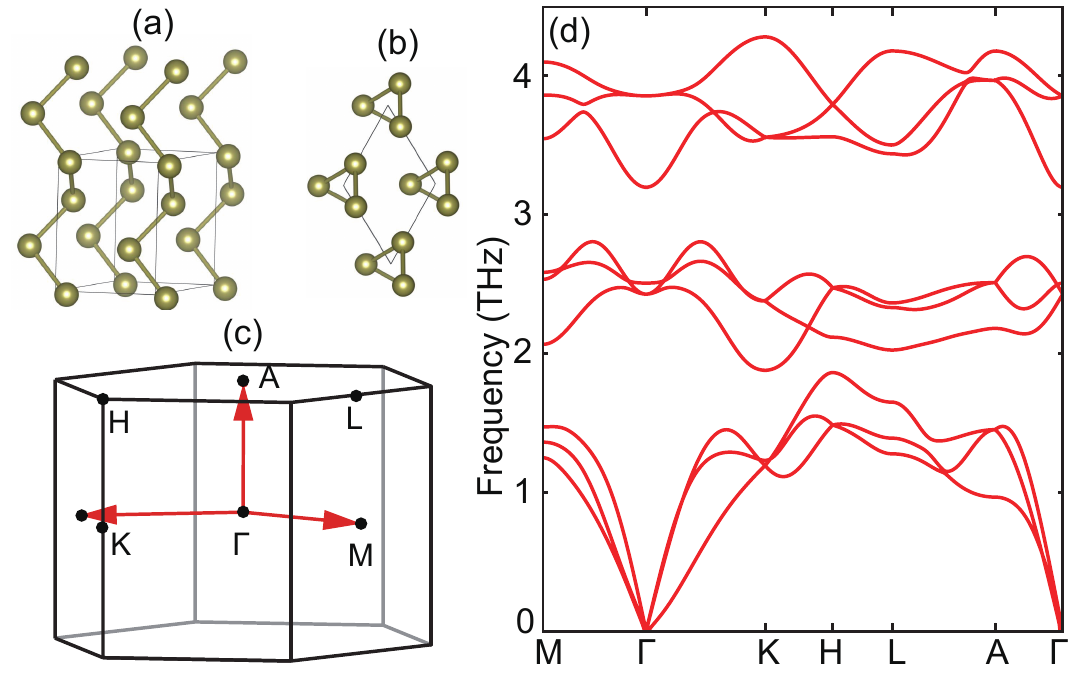}\caption{(a-b) Side and top view of Te, which has a chiral crystal structure with $C_{3}'$ rotation symmetry. (c) Brillouin zone and (d) phonon spectra for Te. 
\label{fig:FIG2}}
\end{figure}
\end{center}

\begin{center}
\begin{figure}
\includegraphics[scale=0.86]{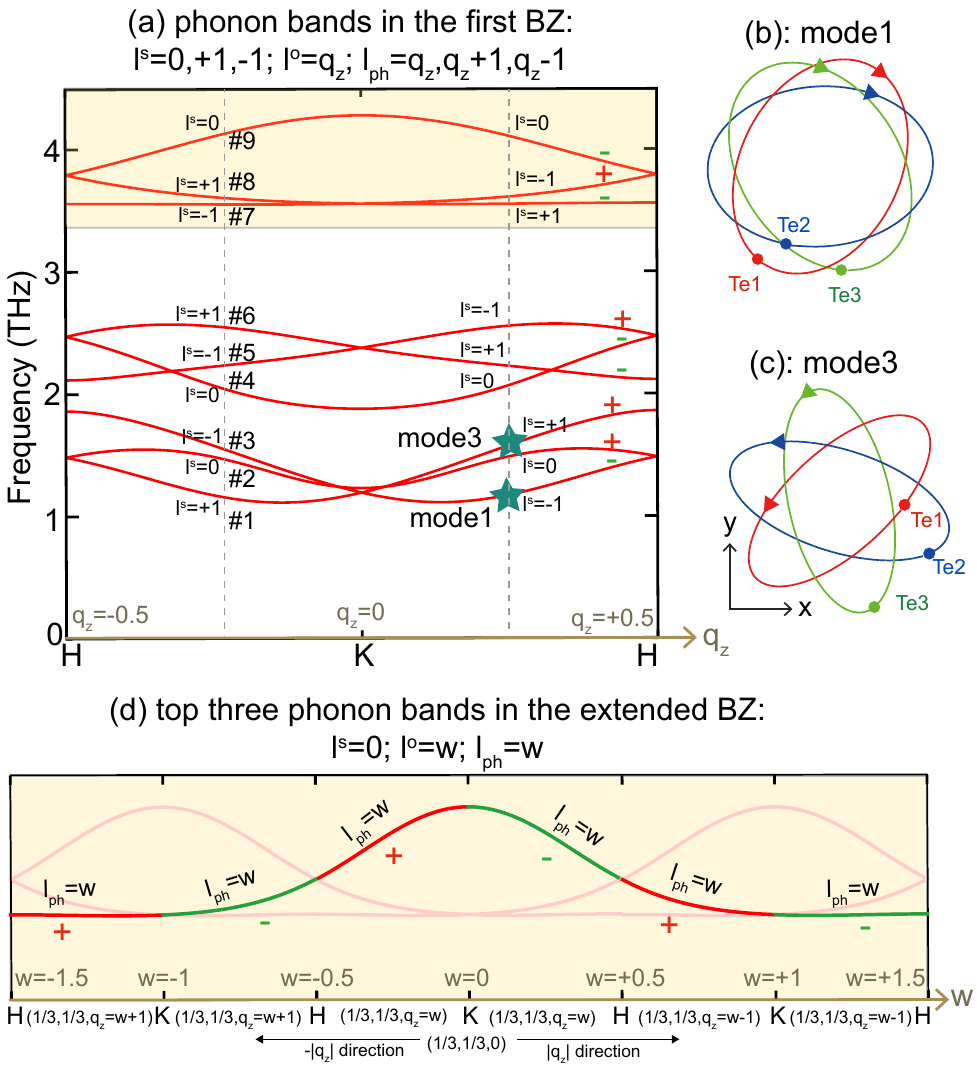}
\caption{(a) Phonon spectra of Te along $H$-$K$-$H$ ($\frac{1}{3},\frac{1}{3},q_{z}$) direction in the first BZ, where $q_{z}\in[-0.5,0.5]$.  ``+'' and ``$-$'' represents  the chirality for each mode and $l^{s}$ is the spin phonon angular momentum for each mode. Both the phonon chirality and $l^{s}$ will keep the same value for each continuous phonon band. 
Real space trajectories for (b) the first and (c) the third mode at ($\frac{1}{3},\frac{1}{3},\frac{1}{4}$) with spin PAM of $l^{s}=-1$ and $l^{s}=+1$, respectively. 
(d)The highest three phonon modes along $H$-$K$-$H$ ($\frac{1}{3},\frac{1}{3},w$) direction crossing three extended BZs, where $l^{s}$=0, $l^{o}$=$l_{ph}$=$w$, and $w\in[-1.5,1.5]$. 
\label{fig:KHmodes}}
\end{figure}
\end{center}

\paragraph*{Crystal and phonon spectra of Te}
Figures \ref{fig:FIG2} (a-b) show the side and top views of Te, which has a chiral crystal structure $P3_{1}21$ (\#152) with a threefold screw rotation $\{C_{3z}'$ = $C_{3z}|(0,0,\frac{1}{3})\}$ along $z$-direction as discussed above. 
Te is a narrow gap semiconductor with a valence band maximum along $K$-$H$ direction, having valleys away from the HSPs. 
Thus, chiral phonons involved in the valley scattering will be also away from HSPs. 
Figure \ref{fig:FIG2} (d) is the phonon spectra of Te along some HSLs, where the names of the HSPs are labeled in Fig. \ref{fig:FIG2} (c). 
Degeneracies at $K$ and $H$ are from the screw rotation symmetry and correspond to topological band crossing with a nonzero Chern number. 


Figure \ref{fig:KHmodes} (a) shows the chirality and $l^{s}$ for each phonon band in the first BZ. For the phonon modes in the first BZ, $l^{s}$ takes the values $0,\pm1$, and it keep the same value for each continuous band, while $l^{o}$=$q_{z}\in[-0.5,0.5]$ is a momentum-dependent quantity. Thus, $l_{ph}$ = $l^{s}$ + $l^{o}$, taking the values $q_{z}$, $q_{z}\pm1$, is also a momentum-dependent quantity. It can be also obtained from the eigenvalue of $\hat{C_{3}^{\prime}}$.

{The behavior of $l_{ph}$ and $l^{s}$ in Fig. \ref{fig:KHmodes} (a) looks complicated, but in fact one can understand it in a simple way as follows. 
In Fig. \ref{fig:KHmodes} (a), nine phonon modes are well separated into three groups, and three phonon modes in each group can be understood as a result of band folding from a single phonon band extending over three BZs. Thus, the values of $l^{s}$, $l^{o}$ and $l_{ph}$ can be understood in an extended scheme over three BZs. 
Figure \ref{fig:KHmodes} (d) shows the top three phonon bands (\#7, \#8 and \#9 in Fig. \ref{fig:KHmodes} (a)) in the extended three BZs, where the horizontal axis is the momentum along $H$-$K$ direction with ($\frac{1}{3},\frac{1}{3},w$) and $w\in[-1.5,1.5]$. In the extended BZ, there will be only one optical phonon band from the vibration of one tellurium, which leads to a trivial value of $l^{s}$: $l^{s}$ = 0. Furthermore, $l^{o}$ = $w$ according to our new definition, so $l_{ph}$ = $l^{s}$ + $l^{o}$ = $w$. 
By noting that the wavenumber $w$ ($-1.5 \leq w \leq +1.5$) along $z$ axis in the extended scheme is related to that in the reduced scheme, $q_{z}$ ($-0.5 \leq q_{z} \leq +0.5$), via $q_{z}$ = $w$ ($mod$ 1), the result in Fig. \ref{fig:KHmodes} (a) immediately follows from Fig. \ref{fig:KHmodes} (d). 
}


In order to show the calculation of $l_{ph}$, $l^{o}$ and $l^{s}$ with our new definition in detail, we will also use Te as an example. 
Table \ref{table:PAM} shows the spin part $l^{s}$ and the orbital part $l^{o}$ of PAM for each phonon mode at ($\frac{1}{3},\frac{1}{3},\frac{1}{4}$), which is the middle momentum of $K$-$H$ line. 
Then, $l^{o}$ is equal to $q_{z}$ for Te as discussed above, thus $l^{o}$=0.25 for each mode of Te. 
Since $l^{s}$ is a trajectory-dependent quantity, we will plot the trajectory for the first and third mode at ($\frac{1}{3},\frac{1}{3},\frac{1}{4}$) to illustrate how to obtain $l^{s}$. Figure \ref{fig:KHmodes} (b-c) show the trajectories of two $C_{3z}^{\prime}$-related atoms for the first and third phonon modes, which is consistent with the $l^{s}$ calculation explained in Fig. \ref{fig:ScrewRot} (e-f).


Table \ref{table:PAM} also shows the results of $l_{ph}$ for all the phonon modes calculated by the two methods: (i) The first one is the summation of the spin and orbital part of PAM, i.e., $l_{ph}$=$l^{o}$+$l^{s}$; (ii) the second one is calculated directly from the eigenvalue of $\hat{C_{3z}^{\prime}}$. The values of $l_{ph}$ calculated by the two methods show a perfect match with each other, which also demonstrate correctness of our definition for phonon PAM $l_{ph}$ in non-symmorphic systems at arbitrary momenta with screw rotation symmetry.

We note that Te has another chiral crystal structure of $P3_{2}21$ (\#154) with a threefold screw rotation $\{C_{3z}'$ = $C_{3z}|(0,0,-\frac{1}{3})\}$ along $z$-direction, which will has the same phonon band spectra with $P3_{1}21$. However, $l_{ph}$, $l^{o}$, and $l^{s}$ will have an additional sign comparing to the ones in $P3_{1}21$ due to the opposite fractional translation along $z$-direction. 
Furthermore, if systems have higher symmetries like chiral cubic symmetry, higher degenerated modes can exist, such as threefold degenerated modes at $\Gamma$, and fourfold degenerated modes at $R (\pi, \pi, \pi)$ for chiral cubic crystals with non-symmorphic symmetries. Likewise, phonon PAM and its orbital/spin part can be defined in the same way.

\begin{figure*}
\includegraphics[scale=0.76]{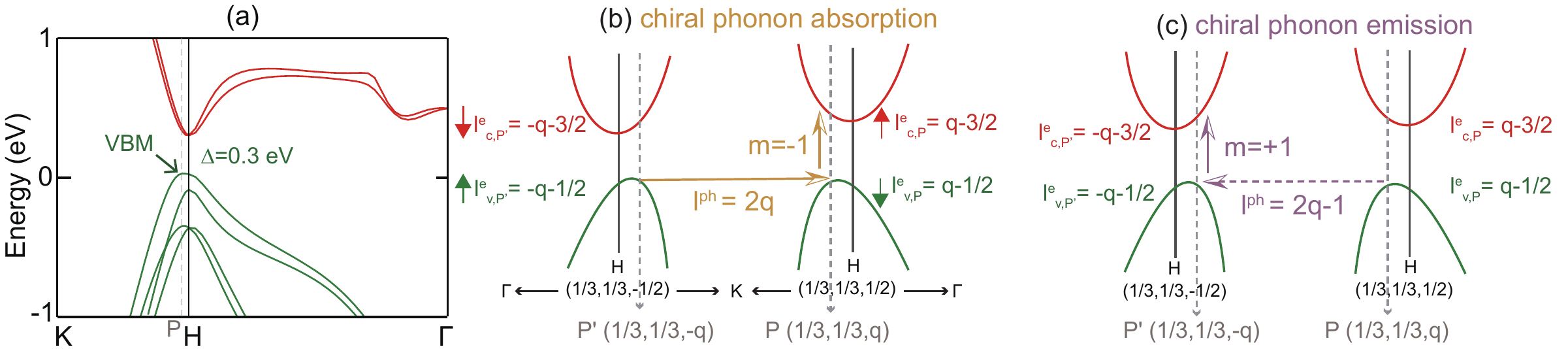}
\caption{New selection rules for phonon-involved intervalley scattering in 3D non-symmorphic crystal Te. (a) Spinful electronic band structure of Te with a direct gap at $P$($\frac{1}{3}$,$\frac{1}{3}$,$q$), which is along $K$-$H$ high-symmetry line and in the vicinity of $H$. (b) By absorbing a chiral phonon with $l^{ph}=2q$, electrons can be scattered between two valleys under the left-circularly polarized light. (c) By emitting a chiral phonon with $l^{ph}=2q-1$, electrons can be scattered between two valleys under the right-circularly polarized light.}
\label{fig:valleys}
\end{figure*}
\twocolumngrid

\begin{table}[]
\begin{tabular}{crccc}
\hline
band order  & $l^{s}$ & $l^{o}$   & $l_{ph}$ calculated by & $l_{ph}$ calculated by   \\ 
&&&$l^{s}+l^{o}$&  eigenvalue of $\hat{C_{3}^{\prime}}$ \\ \hline\hline
\#1 & -1  & 0.25 & -0.75 & -0.75 \\ 
\#2 & 0  & 0.25 & 0.25  &  0.25   \\ 
\#3 & +1 & 0.25 & 1.25  &    1.25   \\ 
\#4 & 0  & 0.25 & 0.25  &     0.25  \\
\#5 & +1 & 0.25 & 1.25  &     1.25  \\ 
\#6 & -1 & 0.25 & -0.75 &     -0.75  \\ 
\#7 & +1 & 0.25 & 1.25  &    1.25   \\ 
\#8 & -1 & 0.25 & -0.75 &     -0.75  \\ 
\#9 & 0  & 0.25 & 0.25  &     0.25  \\ \hline
\end{tabular}
\caption{Orbital part $l^{o}$, spin part $l^{s}$ and total pseudo-angular moment $l_{ph}$ for each phonon band at ($\frac{1}{3},\frac{1}{3},\frac{1}{4}$) in tellurium, which is the middle point of $K$-$H$ line. The results of $l_{ph}$ obtained by two different methods of our new definition match exactly with each other.}
\label{table:PAM}
\end{table}

\paragraph{Electronic band structure of Te}
Figure \ref{fig:valleys} (a) shows the spinful electronic band structure of Te with $P3_{1}21$, by using the crystal lattice constants of $a$ = $b$ = 4.456 $\AA$ and $c$ = 5.592 $\AA$ from experiments \cite{adenis1989reinvestigation}. 
The band structure indicates that Te is a narrow gap insulator with a direct gap about $\Delta$ = 0.3 eV at $P$ ($\frac{1}{3}$,$\frac{1}{3}$,$q$) and $P^{\prime}$ ($\frac{1}{3}$,$\frac{1}{3}$,$-q$), which are along $K$-$H$ direction and in the vicinity of $H$ \cite{hirayama2015weyl}. Due to the large spin-orbit coupling, the highest two valence bands have a very large spin splitting about 0.18 eV, while the lowest two conduction bands have a spin splitting about 30 meV. 
Moreover, the spin component is different for the valance band maximum (VBM) and the conduction band minimum (CBM) at two valleys $P$ and $P^{\prime}$, which forbids the intra-valley transition of electrons at each valley. Thus, both the multi-valley band structure and different spin components for the VBM and CBM bands will be advantageous to realize selective coupling between chiral phonons and valley electrons via photon absorption.

\paragraph{Chiral phonon-involved valley scattering in Te}
As discussed above, the spin conservation condition forbids the intra-valley transition in Te but allows the inter-valley one between the CBM and VBM. In such process, phonons will be involved in the conservations of the total crystal momentum, PAM and energy:
\begin{equation}
\mathbf{q}_{ph}=\mathbf{k}_{e,P}-\mathbf{k}_{e,P^{\prime}},
\end{equation}
\begin{equation}
l_{e,P}^{v}+m+l_{ph}=l_{e,P^{\prime}}^{v}\ mod \ 3,
\label{aa}
\end{equation}
\begin{equation}
h\nu=E_{ph}+\Delta,
\end{equation}
where $m=\pm 1$ represent the right/left-circularly polarized light, $h\nu$ is the energy of the photon, $E_{ph}$ is the energy of the chiral phonon and $\Delta$ is the electronic band gap of Te. 
We note that $m$ is also not a quantized value due to the fractional translation of the screw rotation. However, we take $m=\pm 1$ since the fractional translation of the crystal is negligible compared to the wave length of the photon. 

Figure \ref{fig:valleys} (b) shows the chiral phonon absorption process. 
Since the momentum difference for electrons in the inter-valley transition is $\mathbf{q}_{ph} = \mathbf{k}_{e,P} - \mathbf{k}_{e,P^{\prime}}$ = (0, 0, 2$q$), phonons with momentum of $\mathbf{q}_{ph}$ = (0,0,2$q$) will be involved in this process with possible values of $l_{ph}$ = 2$q$, 2$q$+1, and 2$q$-1. 
As for the PAM conservation, both $l_{e}$ and $l_{ph}$ can be obtained from the eigenvalue of $\hat{C_{3z}^{\prime}}$ at valley $P$ with $q_{z}$ = 2$q$ (or $P^{\prime}$ with $q_{z}$ = $-2q$) by first-principle calculation, thus $l^{e}_{v,P} = q - \frac{1}{2}$, $l^{e}_{v,P^{\prime}} = -q - \frac{1}{2}$, $l^{e}_{c,P} = q - \frac{3}{2}$, $l^{e}_{c,P^{\prime}} = -q - \frac{3}{2}$, and $l^{ph} = 2q$. 
By following Eq. (\ref{aa}), a chiral phonon absorption process can be obtained with $l^{ph} = 2q$ by a left-circularly polarized incident photon with an energy $\Delta$. (Electrons can be also scattered between two valleys by absorbing a chiral phonon with $l^{ph}$ = 2$q$+1 and $m$ = +1.) 
Likewise, the emission process of chiral phonon with $l^{ph} = 2q-1$ can also be detected by a right-circularly polarized light with an energy $\Delta$, which is shown in Fig. \ref{fig:valleys} (c). (Electrons can be also scattered between two valleys by emitting a chiral phonon with $l^{ph}$ = 2$q$ and $m$ = $-$1.) 
In addition, chiral phonons at momentum ($0,0,\pm 2q$) have a nonzero group velocity along $z$-direction, so they can propagate through the material without losing the chirality information when the material thickness is smaller than the phonon scattering length but larger than the electron scattering length.

\paragraph*{Conclusion}
In the previous studies, chiral phonons were studied and observed only at HSPs and in symmorphic space groups, which is mainly restricted by the definition of phonon pseudo-angular momentum including its orbital part and spin part. 
{In this work, we generalize the definition of phonon pseudo-angular momentum (including the orbital part and spin part) to non-symmorphic space groups at arbitrary momenta with screw rotation symmetry, which have a much larger number of systems than the symmorphic ones.  
The redefinition not only makes us realize that phonon pseudo-angular momentum can be non-quantized and $\mathbf{q}$-dependent, but also help us to obtain chiral phonons with a broader condition. For example, Na$_{3}$Sb is a direct gap semiconductor with large Fermi velocity for the valence band maximum along different directions in the vicinity of $A$, which has the same crystal structure with Dirac semimetal Na$_{3}$Bi \cite{wang2012dirac}. 
Since Na$_{3}$Sb preserves $\hat{C_{6}^{\prime}}$ symmetry, non-quantized chiral phonons with PAM $\in$ [-3, 3] $mod$ 6 can be observed by experiments if one gradually modulate the energy of the incident light to excite the electrons along $A$-$\Gamma$ direction (See the appendix for details).  
Moreover, one can explore chiral phonons with a non-vanishing group velocity along any directions, such that the chiral phonons can propagate information in an expected way. }

%
%
%
%
%

\bibliographystyle{unsrt}
\bibliography{reference}

\begin{thebibliography}{10}

\bibitem{nielsen1983adler}
Holger~Bech Nielsen and Masao Ninomiya.
\newblock The {A}dler-{B}ell-{J}ackiw anomaly and {W}eyl fermions in a crystal.
\newblock {\em Physics {L}etters B}, 130(6):389--396, 1983.

\bibitem{son2013chiral}
DT~Son and BZ~Spivak.
\newblock Chiral anomaly and classical negative magnetoresistance of {W}eyl
  metals.
\newblock {\em Physical {R}eview {B}}, 88(10):104412, 2013.

\bibitem{huang2015observation}
Xiaochun Huang, Lingxiao Zhao, Yujia Long, Peipei Wang, Dong Chen, Zhanhai
  Yang, Hui Liang, Mianqi Xue, Hongming Weng, Zhong Fang, et~al.
\newblock Observation of the chiral-anomaly-induced negative magnetoresistance
  in 3{D} {W}eyl semimetal {T}a{A}s.
\newblock {\em Physical {R}eview {X}}, 5(3):031023, 2015.

\bibitem{zhang2005experimental}
Yuanbo Zhang, Yan-Wen Tan, Horst~L Stormer, and Philip Kim.
\newblock Experimental observation of the quantum {H}all effect and {B}erry's
  phase in graphene.
\newblock {\em {N}ature}, 438(7065):201--204, 2005.

\bibitem{zhao2021index}
YX~Zhao and Shengyuan~A Yang.
\newblock Index theorem on chiral {L}andau bands for topological fermions.
\newblock {\em Physical {R}eview {L}etters}, 126(4):046401, 2021.

\bibitem{katsnelson2006chiral}
MI~Katsnelson, KS~Novoselov, and AK~Geim.
\newblock Chiral tunnelling and the {K}lein paradox in graphene.
\newblock {\em Nature {P}hysics}, 2(9):620--625, 2006.

\bibitem{zhang2018double}
Tiantian Zhang, Zhida Song, A~Alexandradinata, Hongming Weng, Chen Fang, Ling
  Lu, and Zhong Fang.
\newblock Double-{W}eyl phonons in transition-metal monosilicides.
\newblock {\em Physical {R}eview {L}etters}, 120(1):016401, 2018.

\bibitem{lu2014topological}
Ling Lu, John~D Joannopoulos, and Marin Solja{\v{c}}i{\'c}.
\newblock Topological photonics.
\newblock {\em Nature Photonics}, 8(11):821, 2014.

\bibitem{lu2015experimental}
Ling Lu, Zhiyu Wang, Dexin Ye, Lixin Ran, Liang Fu, John~D Joannopoulos, and
  Marin Solja{\v{c}}i{\'c}.
\newblock Experimental observation of {W}eyl points.
\newblock {\em Science}, 349(6248):622--624, 2015.

\bibitem{weng2015weyl}
Hongming Weng, Chen Fang, Zhong Fang, B~Andrei Bernevig, and Xi~Dai.
\newblock {W}eyl semimetal phase in noncentrosymmetric transition-metal
  monophosphides.
\newblock {\em Physical {R}eview {X}}, 5(1):011029, 2015.

\bibitem{Weyl_newfermions}
Barry Bradlyn, Jennifer Cano, Zhijun Wang, MG~Vergniory, C~Felser, RJ~Cava, and
  B~Andrei Bernevig.
\newblock Beyond {D}irac and {W}eyl fermions: Unconventional quasiparticles in
  conventional crystals.
\newblock {\em Science}, 353(6299):aaf5037, 2016.

\bibitem{Weyl_Taas}
BQ~Lv, HM~Weng, BB~Fu, XP~Wang, H~Miao, J~Ma, P~Richard, XC~Huang, LX~Zhao,
  GF~Chen, et~al.
\newblock Experimental discovery of {W}eyl semimetal {T}a{A}s.
\newblock {\em Physical {R}eview {X}}, 5(3):031013, 2015.

\bibitem{Weyl_experiments}
Su-Yang Xu, Ilya Belopolski, Nasser Alidoust, Madhab Neupane, Guang Bian,
  Chenglong Zhang, {R}aman Sankar, Guoqing Chang, Zhujun Yuan, Chi-Cheng Lee,
  et~al.
\newblock Discovery of a {W}eyl fermion semimetal and topological fermi arcs.
\newblock {\em Science}, 349(6248):613--617, 2015.

\bibitem{Weyl_exp}
Xiangang Wan, Ari~M Turner, Ashvin Vishwanath, and Sergey~Y Savrasov.
\newblock Topological semimetal and fermi-arc surface states in the electronic
  structure of pyrochlore iridates.
\newblock {\em Physical {R}eview {B}}, 83(20):205101, 2011.

\bibitem{Weyl_phon_mecha}
D~Zeb Rocklin, Bryan Gin-ge Chen, Martin Falk, Vincenzo Vitelli, and
  TC~Lubensky.
\newblock Mechanical {W}eyl modes in topological {M}axwell lattices.
\newblock {\em Physical {R}eview {L}etters}, 116(13):135503, 2016.

\bibitem{Weyl_acoustic3}
Meng Xiao, Wen-Jie Chen, Wen-Yu He, and Che~Ting Chan.
\newblock Synthetic gauge flux and {W}eyl points in acoustic systems.
\newblock {\em Nature Physics}, 11(11):920--924, 2015.

\bibitem{Weyl_acoustic4}
Cheng He, Xu~Ni, Hao Ge, Xiao-Chen Sun, Yan-Bin Chen, Ming-Hui Lu, Xiao-Ping
  Liu, and Yan-Feng Chen.
\newblock Acoustic topological insulator and robust one-way sound transport.
\newblock {\em Nature Physics}, 2016.

\bibitem{rini2007control}
Matteo Rini, Nicky Dean, Jiro Itatani, Yasuhide Tomioka, Yoshinori Tokura,
  Robert~W Schoenlein, Andrea Cavalleri, et~al.
\newblock Control of the electronic phase of a manganite by mode-selective
  vibrational excitation.
\newblock {\em Nature}, 449(7158):72--74, 2007.

\bibitem{forst2015mode}
M~Forst, Roman Mankowsky, and Andrea Cavalleri.
\newblock Mode-selective control of the crystal lattice.
\newblock {\em Accounts of {C}hemical {R}esearch}, 48(2):380--387, 2015.

\bibitem{grissonnanche2020chiral}
G~Grissonnanche, S~Th{\'e}riault, A~Gourgout, M-E Boulanger,
  E~Lefran{\c{c}}ois, A~Ataei, F~Lalibert{\'e}, M~Dion, J-S Zhou, S~Pyon,
  et~al.
\newblock Chiral phonons in the pseudogap phase of cuprates.
\newblock {\em Nature Physics}, 16(11):1108--1111, 2020.

\bibitem{grissonnanche2019giant}
Ga{\"e}l Grissonnanche, Ana{\"e}lle Legros, Sven Badoux, Etienne
  Lefran{\c{c}}ois, Victor Zatko, Maude Lizaire, Francis Lalibert{\'e}, Adrien
  Gourgout, J-S Zhou, Sunseng Pyon, et~al.
\newblock Giant thermal {H}all conductivity in the pseudogap phase of cuprate
  superconductors.
\newblock {\em Nature}, 571(7765):376--380, 2019.

\bibitem{zeng2012valley}
Hualing Zeng, Junfeng Dai, Wang Yao, Di~Xiao, and Xiaodong Cui.
\newblock Valley polarization in {M}o{S}2 monolayers by optical pumping.
\newblock {\em Nature {N}anotechnology}, 7(8):490--493, 2012.

\bibitem{carvalho2017intervalley}
Bruno~R Carvalho, Yuanxi Wang, Sandro Mignuzzi, Debdulal Roy, Mauricio
  Terrones, Cristiano Fantini, Vincent~H Crespi, Leandro~M Malard, and Marcos~A
  Pimenta.
\newblock Intervalley scattering by acoustic phonons in two-dimensional
  {M}o{S}2 revealed by double-resonance {R}aman spectroscopy.
\newblock {\em Nature {C}ommunications}, 8(1):1--8, 2017.

\bibitem{cao2012valley}
Ting Cao, Gang Wang, Wenpeng Han, Huiqi Ye, Chuanrui Zhu, Junren Shi, Qian Niu,
  Pingheng Tan, Enge Wang, Baoli Liu, et~al.
\newblock Valley-selective circular dichroism of monolayer molybdenum
  disulphide.
\newblock {\em Nature {C}ommunications}, 3(1):1--5, 2012.

\bibitem{wang2012electronics}
Qing~Hua Wang, Kourosh Kalantar-Zadeh, Andras Kis, Jonathan~N Coleman, and
  Michael~S Strano.
\newblock Electronics and optoelectronics of two-dimensional transition metal
  dichalcogenides.
\newblock {\em Nature {N}anotechnology}, 7(11):699--712, 2012.

\bibitem{he2014tightly}
Keliang He, Nardeep Kumar, Liang Zhao, Zefang Wang, Kin~Fai Mak, Hui Zhao, and
  Jie Shan.
\newblock Tightly bound excitons in monolayer {WS}e2.
\newblock {\em Physical {R}eview {L}etters}, 113(2):026803, 2014.

\bibitem{wu2013electrical}
Sanfeng Wu, Jason~S Ross, Gui-Bin Liu, Grant Aivazian, Aaron Jones, Zaiyao Fei,
  Wenguang Zhu, Di~Xiao, Wang Yao, David Cobden, et~al.
\newblock Electrical tuning of valley magnetic moment through symmetry control
  in bilayer {M}o{S}2.
\newblock {\em Nature {P}hysics}, 9(3):149--153, 2013.

\bibitem{jones2013optical}
Aaron~M Jones, Hongyi Yu, Nirmal~J Ghimire, Sanfeng Wu, Grant Aivazian, Jason~S
  Ross, Bo~Zhao, Jiaqiang Yan, David~G Mandrus, Di~Xiao, et~al.
\newblock Optical generation of excitonic valley coherence in monolayer {WS}e2.
\newblock {\em Nature {N}anotechnology}, 8(9):634--638, 2013.

\bibitem{xu2014spin}
Xiaodong Xu, Wang Yao, Di~Xiao, and Tony~F Heinz.
\newblock Spin and pseudospins in layered transition metal dichalcogenides.
\newblock {\em Nature Physics}, 10(5):343--350, 2014.

\bibitem{chen2021probing}
Chen Chen, Xiaolong Chen, Bingchen Deng, Kenji Watanabe, Takashi Taniguchi,
  Shengxi Huang, and Fengnian Xia.
\newblock Probing interlayer interaction via chiral phonons in layered
  honeycomb materials.
\newblock {\em Physical {R}eview {B}}, 103(3):035405, 2021.

\bibitem{drapcho2017apparent}
Steven~G Drapcho, Jonghwan Kim, Xiaoping Hong, Chenhao Jin, Sufei Shi,
  Sefaattin Tongay, Junqiao Wu, and Feng Wang.
\newblock Apparent breakdown of {R}aman selection rule at valley exciton
  resonances in monolayer mo s 2.
\newblock {\em Physical {R}eview {B}}, 95(16):165417, 2017.

\bibitem{tatsumi2018interplay}
Yuki Tatsumi and Riichiro Saito.
\newblock Interplay of valley selection and helicity exchange of light in
  {R}aman scattering for graphene and {M}o{S}2.
\newblock {\em Physical {R}eview {B}}, 97(11):115407, 2018.

\bibitem{malard2009raman}
LM~Malard, Marcos~Assun{\c{c}}{\~a}o Pimenta, Gene Dresselhaus, and
  MS~Dresselhaus.
\newblock {R}aman spectroscopy in graphene.
\newblock {\em Physics reports}, 473(5-6):51--87, 2009.

\bibitem{zhu2018observation}
Hanyu Zhu, Jun Yi, Ming-Yang Li, Jun Xiao, Lifa Zhang, Chih-Wen Yang, Robert~A
  Kaindl, Lain-Jong Li, Yuan Wang, and Xiang Zhang.
\newblock Observation of chiral phonons.
\newblock {\em Science}, 359(6375):579--582, 2018.

\bibitem{chen2019entanglement}
Xiaotong Chen, Xin Lu, Sudipta Dubey, Qiang Yao, Sheng Liu, Xingzhi Wang, Qihua
  Xiong, Lifa Zhang, and Ajit Srivastava.
\newblock Entanglement of single-photons and chiral phonons in atomically thin
  {WS}e2.
\newblock {\em Nature Physics}, 15(3):221--227, 2019.

\bibitem{li2019momentum}
Zhipeng Li, Tianmeng Wang, Chenhao Jin, Zhengguang Lu, Zhen Lian, Yuze Meng,
  Mark Blei, Mengnan Gao, Takashi Taniguchi, Kenji Watanabe, et~al.
\newblock Momentum-dark intervalley exciton in monolayer tungsten diselenide
  brightened via chiral phonon.
\newblock {\em {ACS} {N}ano}, 13(12):14107--14113, 2019.

\bibitem{li2019emerging}
Zhipeng Li, Tianmeng Wang, Chenhao Jin, Zhengguang Lu, Zhen Lian, Yuze Meng,
  Mark Blei, Shiyuan Gao, Takashi Taniguchi, Kenji Watanabe, et~al.
\newblock Emerging photoluminescence from the dark-exciton phonon replica in
  monolayer {WS}e2.
\newblock {\em Nature {C}ommunications}, 10(1):1--7, 2019.

\bibitem{zhang2015chiral}
Lifa Zhang and Qian Niu.
\newblock Chiral phonons at high-symmetry points in monolayer hexagonal
  lattices.
\newblock {\em Physical {R}eview {L}etters}, 115(11):115502, 2015.

\bibitem{hanbicki2018double}
Aubrey~T Hanbicki, Hsun-Jen Chuang, Matthew~R Rosenberger, C~Stephen Hellberg,
  Saujan~V Sivaram, Kathleen~M McCreary, Igor~I Mazin, and Berend~T Jonker.
\newblock Double indirect interlayer exciton in a {M}o{S}e2/{WS}e2 van der
  {W}aals heterostructure.
\newblock {\em {ACS} {N}ano}, 12(5):4719--4726, 2018.

\bibitem{kunstmann2018momentum}
Jens Kunstmann, Fabian Mooshammer, Philipp Nagler, Andrey Chaves, Frederick
  Stein, Nicola Paradiso, Gerd Plechinger, Christoph Strunk, Christian
  Sch{\"u}ller, Gotthard Seifert, et~al.
\newblock Momentum-space indirect interlayer excitons in transition-metal
  dichalcogenide van der {W}aals heterostructures.
\newblock {\em Nature Physics}, 14(8):801--805, 2018.

\bibitem{ferreira2010evolution}
EH~Martins Ferreira, Marcus~VO Moutinho, F~Stavale, M{\'a}rcia~Maria Lucchese,
  Rodrigo~B Capaz, Carlos~Alberto Achete, and A~Jorio.
\newblock Evolution of the {R}aman spectra from single-, few-, and many-layer
  graphene with increasing disorder.
\newblock {\em Physical {R}eview {B}}, 82(12):125429, 2010.

\bibitem{hao2010probing}
Yufeng Hao, Yingying Wang, Lei Wang, Zhenhua Ni, Ziqian Wang, Rui Wang,
  Chee~Keong Koo, Zexiang Shen, and John~TL Thong.
\newblock Probing layer number and stacking order of few-layer graphene by
  {R}aman spectroscopy.
\newblock {\em {S}mall}, 6(2):195--200, 2010.

\bibitem{eckmann2012probing}
Axel Eckmann, Alexandre Felten, Artem Mishchenko, Liam Britnell, Ralph Krupke,
  Kostya~S Novoselov, and Cinzia Casiraghi.
\newblock Probing the nature of defects in graphene by {R}aman spectroscopy.
\newblock {\em Nano {L}etters}, 12(8):3925--3930, 2012.

\bibitem{ni2008raman}
Zhenhua Ni, Yingying Wang, Ting Yu, and Zexiang Shen.
\newblock {R}aman spectroscopy and imaging of graphene.
\newblock {\em Nano Research}, 1(4):273--291, 2008.

\bibitem{chen2021propagating}
Hao Chen, Weikang Wu, Jiaojiao Zhu, Shengyuan~A Yang, and Lifa Zhang.
\newblock Propagating chiral phonons in three-dimensional materials.
\newblock {\em Nano {L}etters}, 21(7):3060--3065, 2021.

\bibitem{PhysRevB.92.115202}
First-principles study of anisotropic thermoelectric transport properties of
  {IV-VI} semiconductor compounds {S}n{S}e and {S}n{S}, author = {Guo, Ruiqiang
  and Wang, Xinjiang and Kuang, Youdi and Huang, Baoling}, journal = {Phys.
  Rev. B}.
\newblock 92:115202, Sep 2015.

\bibitem{PhysRevB.94.035304}
Paul~Z. Hanakata, Alexandra Carvalho, David~K. Campbell, and Harold~S. Park.
\newblock Polarization and valley switching in monolayer group-{IV}
  monochalcogenides.
\newblock {\em Phys. Rev. B}, 94:035304, Jul 2016.

\bibitem{PhysRevB.92.085406}
L\'{\i}dia~C. Gomes and A.~Carvalho.
\newblock Phosphorene analogues: Isoelectronic two-dimensional group-{IV}
  monochalcogenides with orthorhombic structure.
\newblock {\em Phys. Rev. B}, 92:085406, Aug 2015.

\bibitem{adenis1989reinvestigation}
CLAIRE Adenis, VRATISLAV Langer, and OLIVER Lindqvist.
\newblock Reinvestigation of the structure of tellurium.
\newblock {\em Acta Crystallographica Section C: Crystal Structure
  {C}ommunications}, 45(6):941--942, 1989.

\bibitem{hirayama2015weyl}
Motoaki Hirayama, Ryo Okugawa, Shoji Ishibashi, Shuichi Murakami, and Takashi
  Miyake.
\newblock {W}eyl node and spin texture in trigonal tellurium and selenium.
\newblock {\em Physical {R}eview {L}etters}, 114(20):206401, 2015.

\bibitem{wang2012dirac}
Zhijun Wang, Yan Sun, Xing-Qiu Chen, Cesare Franchini, Gang Xu, Hongming Weng,
  Xi~Dai, and Zhong Fang.
\newblock Dirac semimetal and topological phase transitions in {A}3{B}i ({A}=
  {N}a, {K}, {R}b).
\newblock {\em Physical {R}eview {B}}, 85(19):195320, 2012.

\end{thebibliography}

\newpage

 \newpage{}
\end{document}